 \documentclass[smallabstract,smallcaptions]{dccpaper}

\usepackage{epsfig}
\usepackage{citesort}
\usepackage{amsmath}
\usepackage{amssymb}
\usepackage{color}
\usepackage{multirow}
\usepackage{booktabs}
\usepackage{url}
\DeclareMathOperator{\EX}{\mathbb{E}}% expected value

\newlength{\figurewidth}
\newlength{\smallfigurewidth}

\begin{document}

\title
{\large
\textbf{End-to-end lossless compression of high precision depth maps guided by pseudo-residual}
}

\author{%
Yuyang Wu$^{\ast, \dag}$, Wei Gao$^{\ast, \dag}\thanks{Corresponding author: Wei Gao (gaowei262@pku.edu.cn). This work was supported in part by Ministry of Science and Technology of China - Science and Technology Innovations 2030 under Grant 2020AAA0103501, in part by Natural Science Foundation of China under Grants 61801303 and 62031013, in part by Guangdong Basic and Applied Basic Research Foundation under Grant 2020A1515012031, in part by Shenzhen Science and Technology Plan Basic Research Project under Grant JCYJ20190808161805519, and in part by Shenzhen Fundamental Research Program under Grant GXWD20201231165807007-20200806163656003.
}$\\[0.5em]
{\small\begin{minipage}{\linewidth}\begin{center}
\begin{tabular}{ccc}
$^{\ast}$School of Electronic and Computer Engineering, Shenzhen Graduate School, \\
Peking University, Shenzhen, China \\
$^{\dag}$ Peng Cheng Laboratory, Shenzhen, China \\
\url{wuyy1234@stu.pku.edu.cn,  gaowei262@pku.edu.cn} 
\end{tabular}
\end{center}\end{minipage}}
}

\maketitle
\thispagestyle{empty}

\begin{abstract}
As a fundamental data format representing spatial information, depth map is widely used in signal processing and computer vision fields. Massive amount of high precision depth maps are produced with the rapid development of equipment like laser scanner or LiDAR. Therefore, it is urgent to explore a new compression method with better compression ratio for high precision depth maps. Utilizing the wide spread deep learning environment, we propose an end-to-end learning-based lossless compression method for high precision depth maps. The whole process is comprised of two sub-processes, named pre-processing of depth maps and deep lossless compression of processed depth maps. The deep lossless compression network consists of two sub-networks, named lossy compression network and lossless compression network. We leverage the concept of pseudo-residual to guide the generation of distribution for residual and avoid introducing context models. Our end-to-end lossless compression network achieves competitive performance over engineered codecs and has low computational cost. 
\end{abstract}

\section{INTRODUCTION}
Depth map is a basic data format in signal processing and computer vision fields. Unlike natural images which represent texture messages, depth maps contain spatial information of physical environments. Therefore, depth maps are widely used in self-driving cars and other 3D scenes. Various kinds of deep learning based work have been done with depth maps including depth estimation \cite{Ranftl2020}, depth completion and so on. These deep learning based work can produce massive amount of depth maps. Besides, as the precision and popularity of equipment like laser scanner or LiDAR increase rapidly, such equipment also produces massive amount of high precision depth maps. Unlike low precision depth maps, high precision depth maps have wider data range and more complex distributions. Facing the storage and 
transmission for large amount of high precision depth maps, it is necessary to create a more efficient compression method. 

Utilizing the wide spread deep learning environments with GPU, we introduce an end-to-end learning-based method to losslessly compress high precision depth maps. Our deep lossless compression network comprises two parts: named lossy compression part and lossless compression part. We introduce the concept of pseudo-residual to help us better predict the distribution of real residual. Generally speaking, our contributions are summarized as follows:
\begin{itemize}
    \item To our knowledge, this is the first end-to-end lossless compression network targeted at the high precision depth maps. 
    \item We leverage pseudo-residual to predict the distribution of real residual without any context modeling, which achieves competitive performance and avoids heavy computational cost.
    \item We conduct experiments on depth maps collected from real world, which have higher precision and more complex distribution than synthetic depth maps. Our method achieves competitive performance over the tested engineered codecs.
    %with laser scanner and LiDAR 
\end{itemize}

\section{Related works}
\label{sec:format}
{\bf Learning-based lossless image compression.} With the rapid development of deep learning, learning-based image compression is impacting the field of conventional image coding. Among various methods, generative model is popular. Generative model is widely used to estimate likelihood and the likelihood is used for entropy coding. The performance of the learning-based compression model depends on how well the estimated distribution approximates the real distribution. The structures of generative models to estimate likelihood are diverse. Variational auto-encoder based generative models are first used in learning-based lossy image compression like \cite{balle2016end}, \cite{cheng2020learned} and \cite{balle2018variational}. Similar structures are adopted for learning-based lossless image compression like \cite{mentzer2019practical}. Auto-regressive based generative models have great performance due to the ability of context modeling. Pixel-CNN and Pixel-RNN \cite{van2016pixel} leverage the context information using a mask convolution operation. \cite{bai2021learning} is the first joint deep lossy and lossless image compression. It first uses a deep lossy compression model to compress the input image and then use a context model to estimate the likelihood of the residual. However, auto-regressive based generative models suffer from heavy computational cost due to context structures. These methods require $O(W \cdot  H)$ forward passes and thus can't be applied to practical usage. \cite{mentzer2020learning}, \cite{mentzer2019practical} and \cite{zhang2020lossless} try to reduce computational cost by removing context models. 

%Recently, flow based generative model also catch attentions for its invertible transformation feature. \cite{van2020idf++} solves the problem of bijective mappings for discrete numbers and applies the invertible flows to lossless compression.

{\bf Non-learning-based depth map compression.} Depth map can be consider as a special format of image. Many well-known engineered codecs provide depth codecs including PNG \cite{boutell1997png}, BPG and FILF \cite{sneyers2016flif}. However, depth map has lots of differences from image. Its data format has one channel while image has three RGB channels. Besides, its content and data range are not the same as image. \cite{morvan2007depth} points out that depth maps contain various smooth areas with sharp edges compared to normal textured images. Therefore, \cite{morvan2007depth} models the depth maps with piecewise-linear function for better lossy compression performance.  Apart from focusing on content features, \cite{lin2011rendering}, \cite{zhang2020fast} and \cite{martinez20133d} propose methods based on data features of depth maps.

\section{METHODS}
\subsection{Overview}
\begin{figure}[h]
\begin{minipage}[b]{1.0\linewidth}
  \centering
  \centerline{\includegraphics[width=14cm]{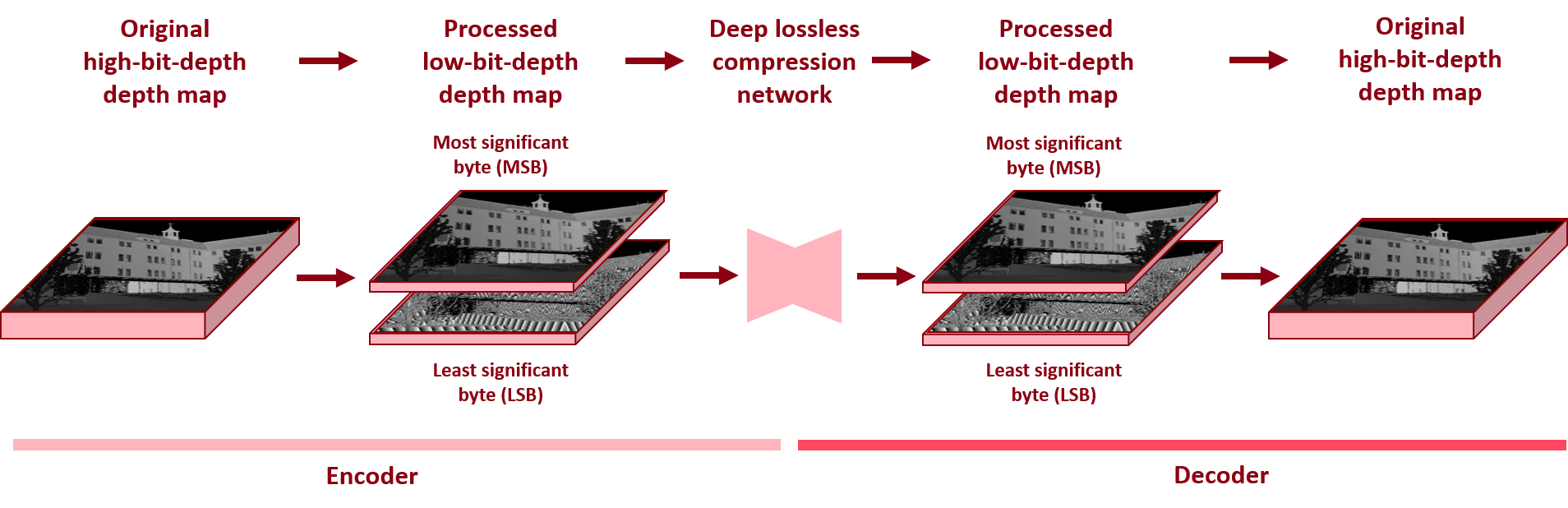}}
\end{minipage}
\caption{Visualization of the whole process. The whole process is comprised of two sub-processes, named pre-processing of depth maps and deep lossless compression of processed depth maps. This example is taken from DIODE dataset.}
\label{fig:whole}
\end{figure}

In this section, we give an overview and describe the details of the end-to-end depth map compression network. The whole process is comprised of two sub-processes, named pre-processing of depth maps and deep lossless compression of processed depth maps. The deep lossless compression network consists of two sub-networks, named lossy compression network and lossless compression network. Details for the pre-processing of depth maps are provided in Section 3.2 and details for the deep lossless compression network are illustrated in Section 3.3.
%The whole compression process is as follows: first, the depth map is encoded and decoded by the lossy compression network. Then the lossy reconstruction of depth map is fed into the second sub-network to produce an estimated distribution of the residuals obtained by the lossy compression. The estimated distribution as well as the residual are sent to the arithmetic coder for lossless compression. Since the residual can be losslessly encoded and decoded, the original depth information can be losslessly preserved. 
The visualization of the whole operations is shown as Figure \ref{fig:whole}.

\subsection{Pre-processing of depth maps}
Since the data range of high precision depth maps is wide, directly inputting depth maps into the end-to-end learning-based compression network could not lead to competitive compression performance. Therefore, we need to preprocess depth maps before compression. We assume that each pixel $x_i$ in a depth map $x=\left \{  x_1, x_2, ..., x_n\right \}$ is an non-negative integer. In high-bit-depth depth maps, the range of $x_i$ can be large such as $0 \leq x_i < 2^{18}$ and so on.  Inspired by \cite{belyaev2017high}, we split the high-bit-depth depth map into two low-bit-depth depth maps. These two low-bit-depth depth maps are denoted as most significant bytes (MSB) and least significant bytes (LSB) respectively. 
Here we define most significant bytes (MSB) as:
\begin{gather}
x_{MSB_i} = \left \lfloor \frac{x_i}{d} \right \rfloor
\end{gather}
and least significant bytes (LSB) as:
\begin{gather}
x_{LSB_i} = x_i \bmod d
\end{gather}
where $d$ is a predefined arbitrary integer like 256, 128, 64 and so on.

After splitting, MSB and LSB are concatenated and sent to lossless compression network. With losslessly decoding, the original depth map can be recovered by:
\begin{gather}
x_i = x_{MSB_i}*d+x_{LSB_i} 
\end{gather}

\subsection{Joint lossy depth maps \& lossless residual compression}

\begin{figure*}
    \centering
    \includegraphics[scale=0.45]{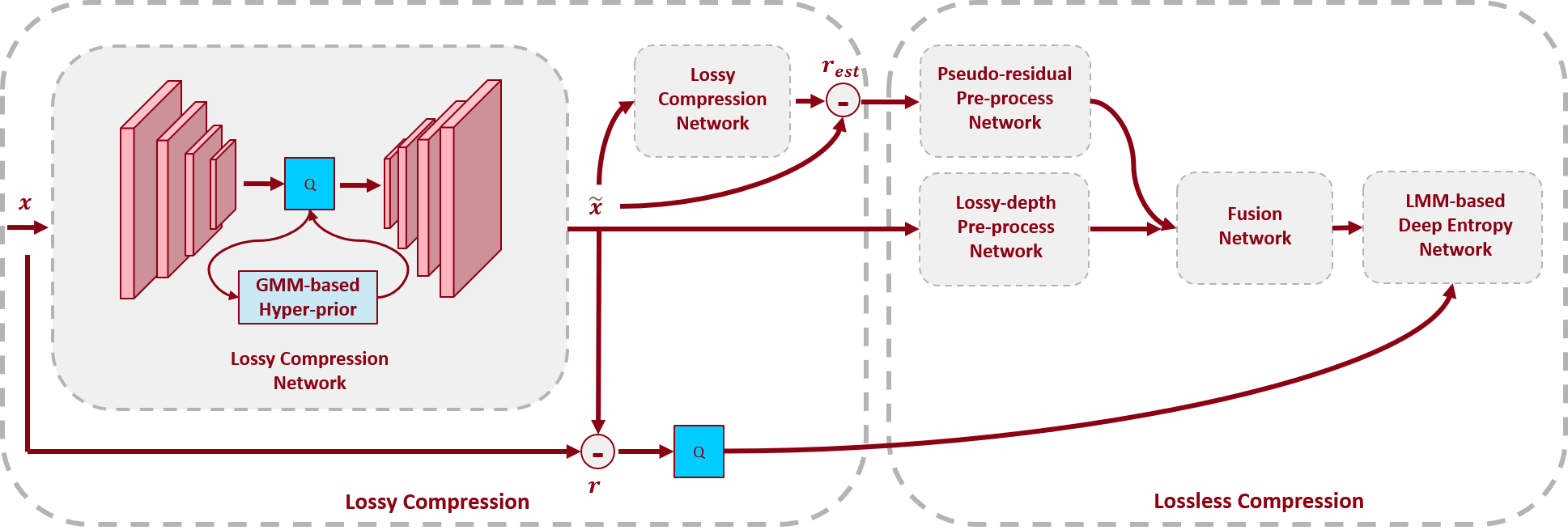}
    \caption{The overview of the network. The whole network comprises two parts, named lossy compression and lossless compression. We leverage pseudo-residual to guide the distribution modeling for real residual. $Q$ denotes quantization.}
    \label{fig:network_overview}
\end{figure*}

The overview of the network is shown in Figure \ref{fig:network_overview}. The whole network comprises two parts, named lossy compression and lossless compression. The input of lossy compression $x$ is the concatenation of $x_{MSB}$ and $x_{LSB}$. The structure of lossy compression network is similar to \cite{cheng2020learned} but without any context models. Lossy compression network produces a lossy reconstruction depth map named $\tilde x$. The lossy reconstruction $\tilde x$ is fed into the lossy compression network again to obtain $\tilde x_{sim}$. Note that we use lossy compression network twice here for different purposes. The first usage is to obtain a real residual denoted as $r$ and the second usage is to obtain an estimated pseudo-residual denoted as $r_{est}$. $r_{est}$ is an estimated pseudo-residual used to guide the generation of distribution for $r$. We define the operation of lossy compression network as $C()$ and the whole process described above is as follows:
\begin{gather}
	\tilde x=C(x) \\
    \tilde x_{sim}=C(\tilde x) \\
    r_{est} = \tilde x - \tilde x_{sim}
\end{gather}
Different from the definition of pseudo-residual $r_{est}$, real residual $r$ is obtained as:
\begin{gather}
	r = \mbox{round}(x-\tilde x)
\end{gather}
Note that since the operation $\mbox{round}$ is not differentiable, we add uniform noise instead during training and use operation $\mbox{round}$ for inference.

So far we have $\tilde x$, $r_{est}$ and $r$. $\tilde x$ and $r_{est}$ are input into the lossless residual network to generate an estimated distribution of $r$ which we can use to losslessly encode and decode $r$ using an arithmetic coder.

\begin{figure}[t]
    \begin{center}
    \begin{tabular}{cc}
    \includegraphics[width=6cm]{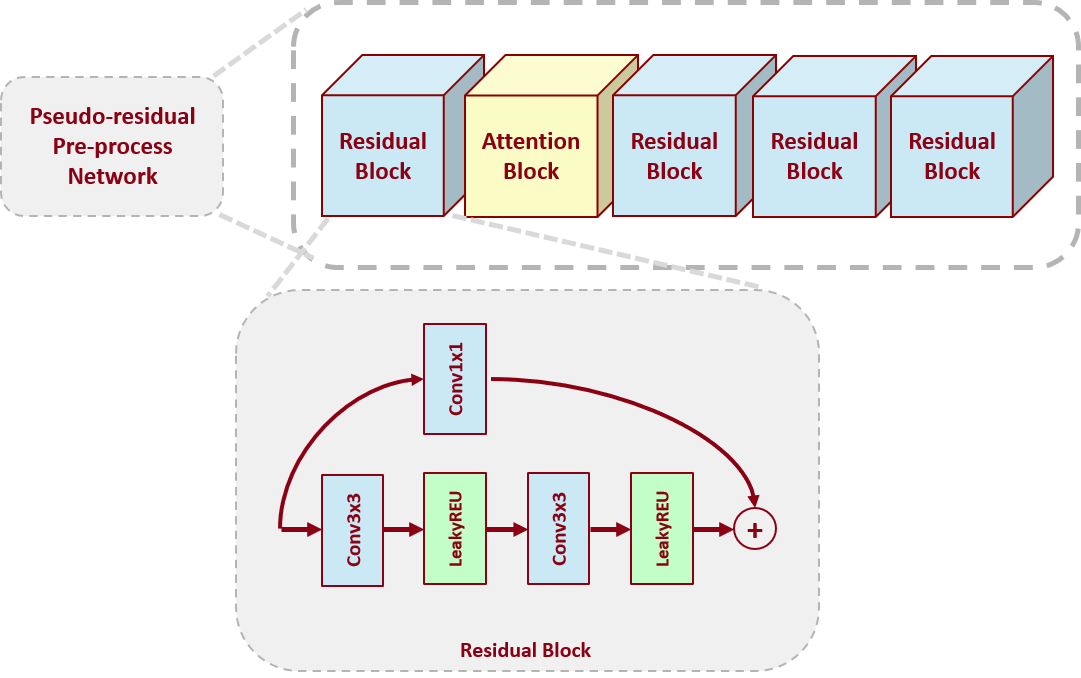} &
    \includegraphics[width=9cm]{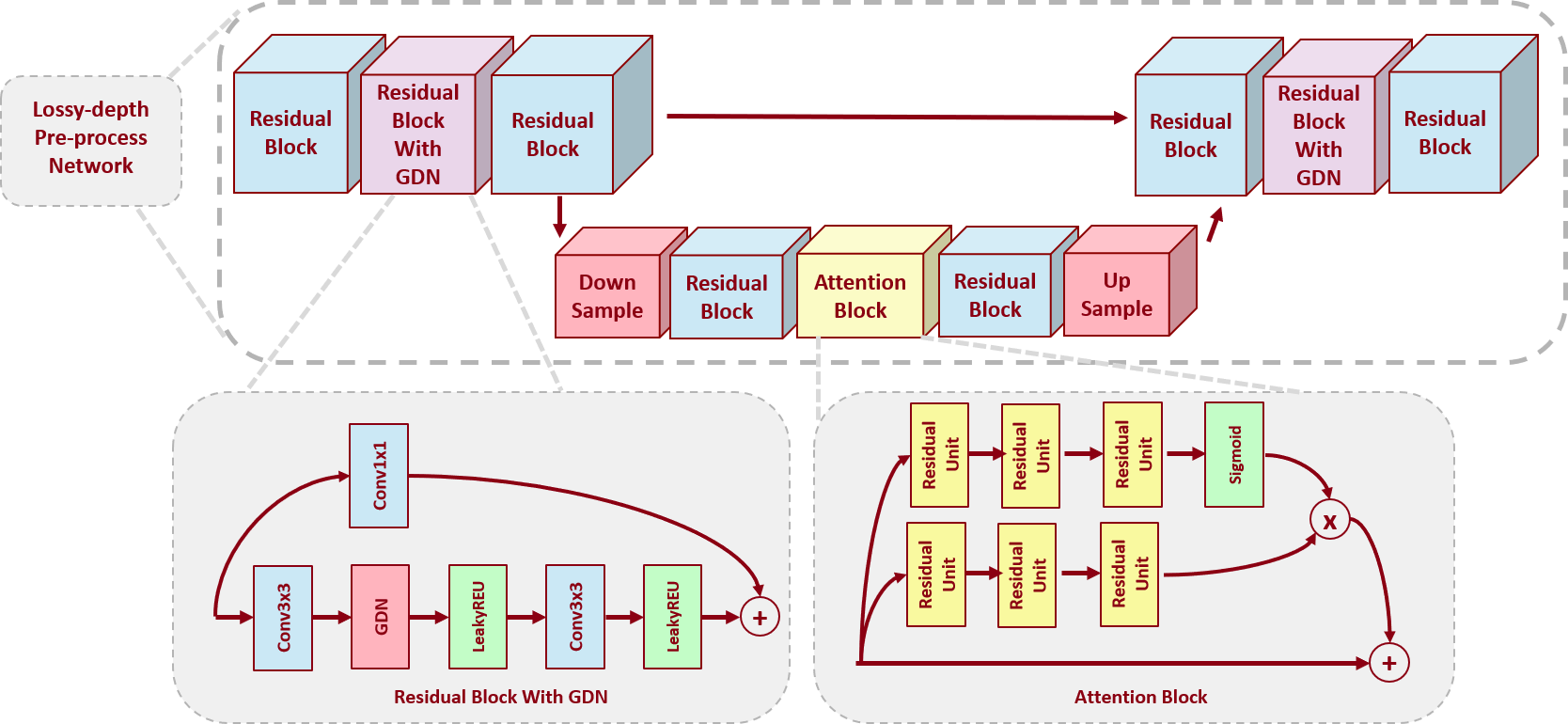} \\
    {\small (a) }  & {\small (b)}
    \end{tabular}
    \end{center}
    \caption{\label{fig:preprocess_network} Details of pseudo-residual pre-process network and lossy-depth pre-process network. These two networks pre-process $r_{est}$ and $\tilde x$, respectively.  }
\end{figure}

$r_{est}$ is fed into the pseudo-residual pre-process network and $\tilde x$ is fed into the lossy-depth pre-process network. As shown in Figure \ref{fig:preprocess_network}, pseudo-residual pre-process network comprises four residual blocks and one attention block. The architecture of lossy-depth pre-process network is more complex than the pseudo-residual pre-process network since the nonlinear transformation for $\tilde x$ is more complicated. Lossy-depth pre-process network is an U-net like network. We down-sample the feature to study features from different scales as well as reduce the computational cost. Down-sample operation is a 3x3 convolution with stride 2 and up-sample operation is a 3x3 sub-pixel convolution. After the up-sample operation, features from lower scale are concatenated with the features from skip connection. Residual blocks in lossy-depth pre-process network are the same as the ones in pseudo-residual pre-process network. Structures of other blocks are illustrated in Figure \ref{fig:preprocess_network}.

\begin{figure}[h]
  \centering
  \centerline{\includegraphics[width=13cm]{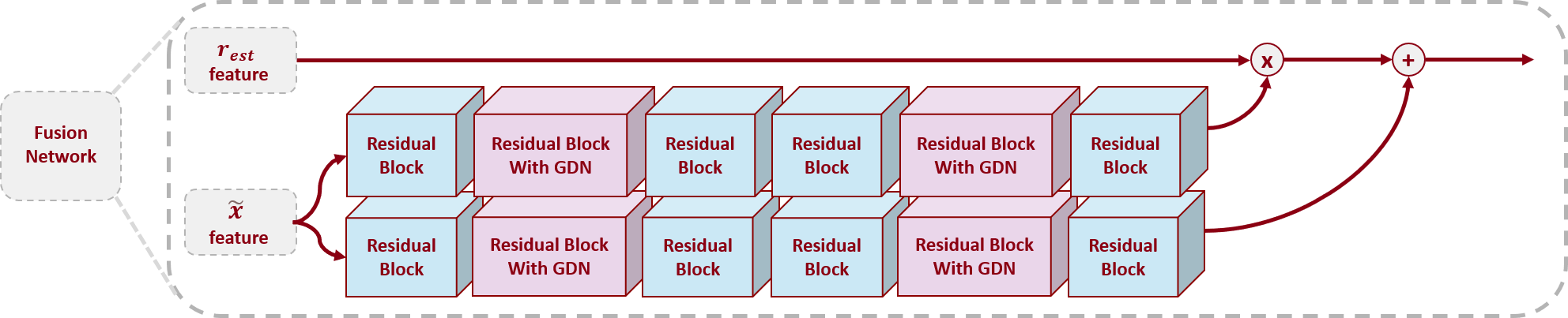}}
\caption{Details of fusion network. Features from pseudo-residual pre-process network and lossy-depth pre-process network are fused with the fusion network.}
\label{fig:fusion_network}
\end{figure}
Output features from pseudo-residual pre-process network and lossy-depth pre-process network are further fused with a fusion network. Details of fusion network are shown in Figure \ref{fig:fusion_network}. The structures of blocks in the fusion network are the same as the ones in lossy-depth pre-process network. With operations of multiplication and addition, features from $\tilde x$ and $r_{est}$ are fused together and are sent to the last network named LMM-based deep entropy network.

\begin{figure}[h]
  \centering
  \centerline{\includegraphics[width=12cm]{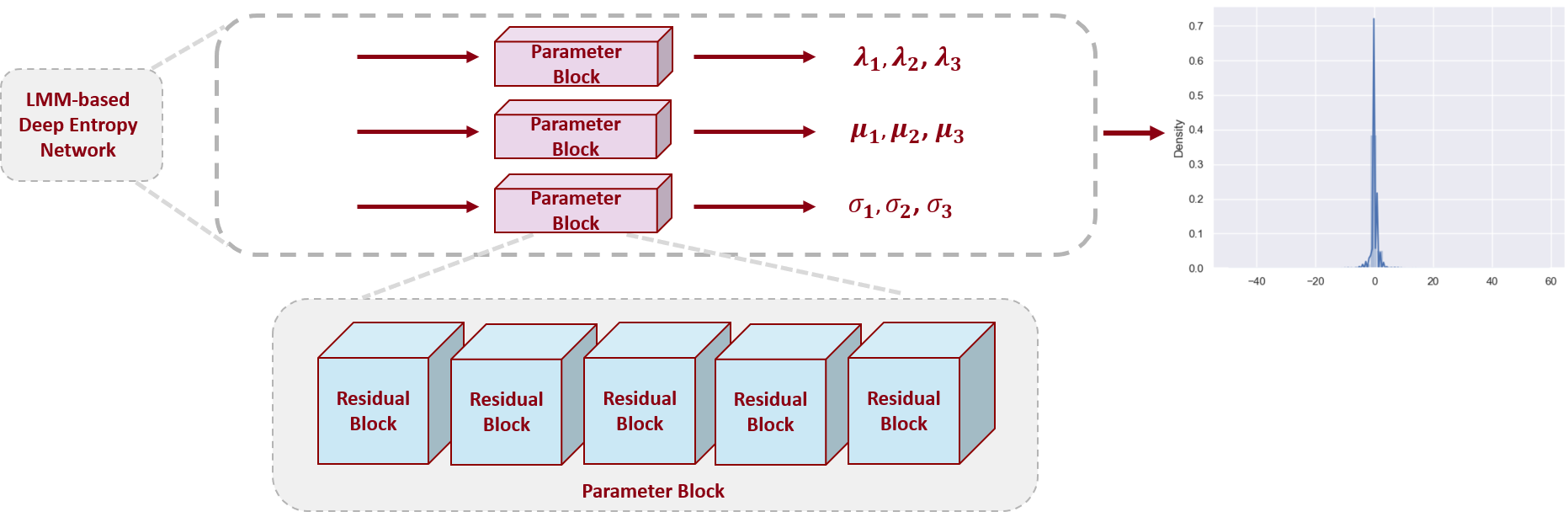}}
\caption{Details of LMM-based deep entropy network. LMM means Laplace mixture distribution. The parameters of Laplace mixture distribution are outputs of three parameter blocks.}
\label{fig:entropy_network}
\end{figure}

We model the distribution of residual $r$ with a Laplace mixture distribution. For a single Laplace distribution, we have:

\begin{gather}
	f(r|\sigma, \mu) = \frac{1}{{2\sigma }}{e^{ - \frac{{\left| {r   - \mu } \right|}}{\sigma }}} 
\end{gather}
Laplace mixture distribution is defined as:
\begin{gather}
    \label{Equ:LMM}
	P\left( {{r}|\lambda,\sigma, \mu } \right) = \sum\limits_{k = 1}^K {{\lambda_k}f\left( {{r}|{\sigma _k},{\mu _k}} \right)},
	with \sum\limits_{k = 1}^K {\lambda_k} =1
\end{gather}
In this work, we set $K=3$. The parameters of Laplace mixture distribution are outputs of three parameter blocks. Each parameter block consists of five residual blocks. Residual blocks are the same as the ones in pseudo-residual pre-process network.

\subsection{Loss function}
Based on Shannon entropy \cite{shannon2001mathematical}, the entropy of $p$ is equal to the expected number of bits needed to encode each $x_i$:
\begin{gather}
	H(p) = bits(x_i) = \EX_{x_i \sim \tilde p}[-\log_2 p(x_i)]
\end{gather}
where $\tilde p$ is the real distribution of $x$ and $p$ is the estimated distribution generated by our model.
Overall, there are two compression processes: the first one is a lossy compression and the other is a lossless compression. Therefore, the whole compression rate consists of two parts, named $R_{lossy}$ and $R_{lossless}$. Since we adopt a hyper prior based structure for lossy compression, there are two features needed to compress named latent feature $\hat y$ and hyper prior feature $\hat z$. Thereby, $R_{lossy}$ contains two parts:
\begin{gather}
	R_{lossy} = \EX_{\tilde p(x)}\EX_{q(\hat y, \hat z|x)}[-\log_2 p(\hat z)-\log_2 p(\hat y|\hat z)]
\end{gather}
while $R_{lossless}$ is defined as:
\begin{gather}
	R_{lossless} = \EX_{\tilde p(x)}\EX_{q(\tilde x, r_{est}|x)}[-\log_2 p( r |\tilde x, r_{est})]
\end{gather}
where $r$ is the real residual, $r_{est}$ is the estimated pseudo-residual and $\tilde x$ is the lossy reconstruction of input depth map $x$. The definitions of these variables are given in Section 3.3. 

Based on the Laplace mixture distribution in eq. (\ref{Equ:LMM}), we evaluate rate terms at discrete $r$ via its CDF following \cite{mentzer2019practical} and \cite{mentzer2020learning}:
\begin{gather}
	p(r) = \mbox{CDF}(r+ \frac{1}{2}) - \mbox{CDF}(r- \frac{1}{2}) 
\end{gather}

Besides rate terms $R_{lossy}$ and $R_{lossless}$, we add two distortion terms:
\begin{gather}
D(x, \tilde x) = \EX_{\tilde p(x)}\EX_{q(\tilde x|x)}||x-\tilde x ||_2\\
D(r, r_{est}) = \EX_{\tilde p(x)}\EX_{q(r, r_{est}|x)}||r- r_{est} ||_2
\end{gather}
where $D(x, \tilde x)$ aims to minimize the mean square error (MSE) between the raw input $x$ and its lossy reconstruction $\tilde x$; $D(r, r_{est})$ is used to minimize the gap between real residual $r$ and its estimated version pseudo-residual $r_{est}$.

The full joint lossy depth map \& lossless residual compression is optimized with:
\begin{gather}
	L = R_{lossy} + R_{lossless} + \alpha * D(x, \tilde x)  + \beta * D(r, r_{est})
\end{gather}
where $\alpha$ and $\beta$ are two hyper parameters that control the reconstruction performance of lossy depth and generation performance of  pseudo-residual, respectively.  

\section{Experiment results}

\subsection{Datasets}
\begin{itemize}
    \item {\bf DIODE} (Dense Indoor/Outdoor DEpth) \cite{vasiljevic2019diode} is the first public dataset that includes RGBD images from both indoor and outdoor scenes. It contains thousands of accurate, dense and long-range depth maps. This dataset is collected by a FARO Focus S350 scanner. The FARO is a laser scanner that provides highly accurate depth measurements ranging from 0.6m to 350m. A depth map of DIODE contains two parts, one is the raw depth obtained by the sensor and the other is the valid mask. Mask is used to mark whether a pixel is valid or not. In our experiment, we set the pixels with invalid mask in raw depths as zero. Since the test dataset of DIODE is not released, we split the original training dataset into our training dataset and validation dataset: we use $95\%$ of original training dataset as our training dataset and the rest $5\%$ as our validation dataset. We set the original validation dataset as our test dataset. Our test dataset of DIODE contains 771 depth maps.
    \item {\bf SementicKITTI } \cite{behley2019semantickitti} contains 22 point cloud sequences obtained by a Velodyne HDL-64E LiDAR. Light detection and ranging (LiDAR) provides precise geometric information about the environment. We convert the point cloud data into depth map same as range image representation. The depths range from 0.8m to 120m. We use median filter to process missing pixels. We use sequences $00\sim 10$ (except 08) as our training dataset and 08 as the validation dataset. We use sequences $11\sim 21$ as the test dataset. Our test dataset of SementicKITTI contains 20351 depth maps.
\end{itemize}

\subsection{Training and evaluation}

The  whole  network  comprises two  parts,  named  lossy  compression  and  lossless  compression. We set the channel numbers of networks from lossy  compression part as 196 and set the channel numbers of networks from lossless  compression part as 64. 

For hyper parameters, we set $\alpha$ and $\beta$ as 25. We set the initial learning rate as 0.00015 and learning rate decays every 20 epochs by a factor of 0.75 for 200 epochs. We use random $256\times 64$ crops extracted from the training dataset. We use Adam as our optimizer and parameters $\beta_1$ and $\beta_2$ are set to 0.9, 0.999, respectively. The batch size is set to 16. 

\subsection{Lossless compression performance comparison}
 
\begin{table}[tp]
\begin{center}
\caption{\label{tab:bpp_result}%
Compression performance comparison of our method and seven engineered codecs. We show lossless compression performance in terms of bpp (bits per pixel). The precision of depth maps is $1mm$.}
{
\renewcommand{\baselinestretch}{1}\footnotesize
\begin{tabular}{|c|c|c|c|}
\hline
\multicolumn{2}{|c|}{coding schemes}& DIODE  & SementicKITTI \\
\hline
\multirow{4}{*}{non-learned data lossless}   & ZLIB & 8.4090 &    13.4663   \\
                           & GZIP & 7.9963 &    12.8679    \\
                           & BZ2  & 4.9614 &     9.7833  \\
                           & LZMA & 4.9542 &     8.5063   \\
                           \hline
\multirow{5}{*}{non-learned image lossless}  & AVIF & 6.3712 &    11.9195    \\
                           & BPG  & 5.5042 &     10.0245   \\
                           & PNG  & 5.2862 &    10.0981    \\
                           & FILF & 4.0304 &    10.8983    \\
                           & WEBP & 4.0807 &     7.9249   \\
                           \hline
\multirow{2}{*}{learned image lossless  }   & IDF  & 5.4870 &  9.7110  \\
                           & SReC & 4.9668 & 8.9721 \\
                           \hline
learned depth map lossless&  Ours    & \bf{3.7514} & \bf{7.3725}   \\
\hline
\end{tabular}}
\end{center}
\end{table}

We compare our proposed method with nine non-learned codecs and two learned codecs, including lossless data compression methods and lossless image compression methods. For non-learned lossless data compression methods, we use  ZLIB, GZIP, BZ2 and LZMA. For non-learned lossless image compression methods, we use BPG, PNG, AVIF, WEBP and FILF. For learned lossless image compression methods, we retrain two popular methods, named SReC\footnote{SReC: https://github.com/caoscott/SReC} \cite{cao2020lossless} and IDF\footnote{ IDF: https://github.com/jornpeters/integer\_discrete\_flows} \cite{hoogeboom2019integer}, based on their official public implementations. For fairness, we set the optimization strategies exactly the same for all learned methods, including the same input, same epochs, same learning rate, same learning rate decay strategy and same optimizer. 

We set the precision of depth maps as $1mm$ and all depth maps are scaled and quantized before sent to the coder. For image based methods, we losslessly convert the original high precision depths map into a three channel 8-bit-depth RGB image as input. All experiment is based on the open source implementations. We show lossless compression performance in terms of  bpp (bits per pixel). Detailed experiment results are shown in Table. \ref{tab:bpp_result}. Results prove that our proposed method achieves competitive performance. SReC and IDF fail to achieve good performance partly due to the fact that their models are designed based on the distribution of normal RGB images, which may not work for high precision depth maps.

\subsection{Sensitivity analysis}

Since we have pre-processing of high precision depth maps as described in Section 3.2, we conduct sensitivity analysis of hyper parameter $d$ on both DIODE and SementicKITTI. We report the bpp of $R_{lossy}$, $R_{lossless}$ and overall bpp. The definition of $R_{lossy}$ and $R_{lossless}$ are given in Section 3.4. Results show that $d$ has slight impact on the $R_{lossy}$, $R_{lossless}$ and overall bpp. Detailed experiment results are shown in Table. \ref{tab:bpp_sensitivity}. Besides, we also compare the performance using different hyper parameters $\alpha$ and $\beta$. The results in Table. \ref{tab:bpp_sensitivity2} show that decreasing the  hyper parameters $\alpha$ and $\beta$ leads to better compression rate.

\begin{table}[tp]
\begin{center}
\caption{\label{tab:bpp_sensitivity}%
Sensitivity analysis for hyper parameter $d$ with $\alpha=100$ and $\beta=100$.  }
{
\renewcommand{\baselinestretch}{1}\footnotesize
\begin{tabular}{|c|c|c|c|c|c|}
\hline
\multirow{2}{*}{datasets}  & \multirow{2}{*}{hyper parameter setting}  &  \multicolumn{2}{|c|}{$R_{lossy}$} &  \multirow{2}{*}{$R_{lossless}$} & \multirow{2}{*}{overall bpp} \\
\cline{3-4}
 & & $R_{\hat{y}}$ & $R_{\hat{z}}$ & & \\
\hline
\multirow{5}{*}{DIODE} &   $d=1024$  & 0.3630 & 0.0100 & 3.5692 &  3.9423  \\
     &   $d=512$  & 0.3413 & 0.0101 &  3.4838  & \bf{3.8352}  \\
     &   $d=256$  & 0.3590 & 0.0103 &  3.4949 & 3.8642   \\
     &   $d=128$   & 0.3573& 0.0103 &   3.5680 & 3.9357      \\
      & $d=64$   & 0.3625 & 0.0099 &    3.6571& 4.0295   \\
                           \hline
\multirow{5}{*}{SementicKITTI} & $d=128$ & 0.3734 & 0.0090 & 7.2628 &7.6453   \\
             & $d=64$    &0.3683 & 0.0088 &  7.3465& 7.7236 \\
             &   $d=32$   & 0.3813&0.0093 & 7.1916 & 7.5823    \\
             &   $d=16$   & 0.4245&  0.0096& 7.0674&  \bf{7.5015}  \\
             &   $d=8$    & 0.4913 &  0.0099  &  7.0735&  7.5747  \\
             \hline
\end{tabular}}
\end{center}
\end{table}

\begin{table}[tp]
\begin{center}
\caption{\label{tab:bpp_sensitivity2}%
Sensitivity analysis for hyper parameters $\alpha$ and $\beta$.  }
{
\renewcommand{\baselinestretch}{1}\footnotesize
\begin{tabular}{|c|c|c|c|c|c|}
\hline
\multirow{2}{*}{datasets}  & \multirow{2}{*}{hyper parameters setting}  &  \multicolumn{2}{|c|}{$R_{lossy}$} &  \multirow{2}{*}{$R_{lossless}$} & \multirow{2}{*}{overall bpp} \\
\cline{3-4}
 & & $R_{\hat{y}}$ & $R_{\hat{z}}$ & & \\
\hline
\multirow{5}{*}{DIODE} &   $\alpha=400$, $\beta=400$, $d=512$  & 0.3681 & 0.0102 & 3.5198 &  3.8981  \\
     &    $\alpha=200$, $\beta=200$, $d=512$  & 0.3493 & 0.0102 &  3.4500  & 3.8096  \\
     &   $\alpha=100$, $\beta=100$, $d=512$ & 0.3413 & 0.0101 &  3.4838 & 3.8352   \\
     &   $\alpha=50$, $\beta=50$, $d=512$  & 0.3316& 0.0103 &    3.4251 & 3.7669      \\
      & $\alpha=25$, $\beta=25$, $d=512$  &  0.3265 &  0.0103 &   3.4147& \bf{3.7514}   \\
\hline
\multirow{5}{*}{SementicKITTI} &   $\alpha=400$, $\beta=400$, $d=16$  &  0.4840 & 0.0100& 7.1261 & 7.6201\\
     &    $\alpha=200$, $\beta=200$, $d=16$  &0.4533 &0.0099 & 7.0739  & 7.5370 \\
     &   $\alpha=100$, $\beta=100$, $d=16$  &0.4245 & 0.0096&7.0674  & 7.5015 \\
     &   $\alpha=50$, $\beta=50$, $d=512$  $d=16$  &0.4112 &0.0097& 7.0195  &  7.4405  \\
      & $\alpha=25$, $\beta=25$, $d=512$  $d=16$  & 0.4004 &0.0097& 6.9624  & \bf{7.3725}   \\
\hline
\end{tabular}}
\end{center}
\end{table}

\subsection{Ablation study}
In order to prove the effectiveness of the proposed network architecture, we give an ablation study on two modules, named fusion network and LMM-based deep entropy network. We replace fusion network with a simple operation of concatenation. For LMM-based deep entropy network, we replace the Laplace mixture distribution with Logistic mixture distribution. Since directly removing pseudo-residual modules leads to unstable training process, we do not include that part into the ablation study. Detailed results are given in Table. \ref{tab:ablation_study}. 

\begin{table}[tp]
\begin{center}
\caption{\label{tab:ablation_study}%
Ablation study on fusion network and LMM-based deep entropy network with $\alpha=25$, $\beta=25$, $d=512$.  }
{
\renewcommand{\baselinestretch}{1}\footnotesize
\begin{tabular}{|c|c|c|c|c|c|c|}
\hline
\multirow{3}{*}{datasets}  & \multicolumn{2}{|c|}{modules}  &  \multicolumn{2}{|c|}{$R_{lossy}$} &  \multirow{3}{*}{$R_{lossless}$} & \multirow{3}{*}{overall bpp} \\
\cline{2-5}
 & \multirow{2}{*}{fusion network} & LMM-based deep  & \multirow{2}{*}{$R_{\hat{y}}$ }& \multirow{2}{*}{$R_{\hat{z}}$} & & \\
  &  & entropy network &  &  & & \\
\hline
\multirow{4}{*}{DIODE} &   \checkmark &\checkmark  &0.3265 &  0.0103 &   3.4147& \bf{3.7514}   \\
     &    & \checkmark & 0.3002 & 0.0101 & 3.5955 & 3.9059  \\
     &  \checkmark &  & 0.3322 & 0.0103 & 3.5058&  3.8483 \\
     &    &  & 0.2516 & 0.0093 & 3.7428 &  4.0037   \\
\hline
\end{tabular}}
\end{center}
\end{table}

\subsection{Computational cost}
Although our network guided by pseudo-residual requires a two-time feed-forward inference, we do not include any context models in our network. Without context models, we need only $O(1)$ computational cost other than $O(W \cdot  H)$ computational cost. We run tests with a Nvidia GeForce RTX 2080 Ti GPU and four Intel Xeon Gold 6148 CPUs. We set input size as $256\times 64$ with minibatch size as 1. Each computation costs around 0.04s.

\section{Conclusion}

Facing the explosion of massive amount of high precision depth maps, we propose a novel end-to-end lossless compression method for high precision depth maps. The whole process is comprised of two sub-processes, named pre-processing of depth maps and deep lossless compression of processed depth maps. The deep lossless compression network consists of two sub-networks, named lossy compression network and lossless compression network. We utilize the concept of pseudo-residual to guide the generation of distribution for residual. Our computational cost is low since we do not use context models. We conduct experiments on datasets collected by high precision laser scanner or LiDAR. Experiment results prove the effectiveness of our end-to-end lossless compression. To further improve our method, we could investigate the weak form of context information. Besides, domain transfer and network self-adaptation remain to explore to improve robustness for practical usage. 

\Section{References}
\bibliographystyle{IEEEbib}
\bibliography{refs}

\end{document}